\documentclass[aip,pop,reprint]{revtex4-1}
\usepackage{graphicx}
\usepackage[breaklinks=true,colorlinks=true,linkcolor=blue,urlcolor=blue,citecolor=blue]{hyperref}

\newcommand{\vc}{\mathbf}

\newsavebox{\myimage}

\draft % marks overfull lines with a black rule on the right

\begin{document}

% Use the \preprint command to place your local institutional report number 
% on the title page in preprint mode.
% Multiple \preprint commands are allowed.
%\preprint{}

\title{Particle-in-cell study of the ion-to-electron sheath transition} %Title of paper

% repeat the \author .. \affiliation  etc. as needed
% \email, \thanks, \homepage, \altaffiliation all apply to the current author.
% Explanatory text should go in the []'s, 
% actual e-mail address or url should go in the {}'s for \email and \homepage.
% Please use the appropriate macro for the type of information

% \affiliation command applies to all authors since the last \affiliation command. 
% The \affiliation command should follow the other information.

\author{Brett Scheiner}
\email[]{brett-scheiner@uiowa.edu}
%\homepage[]{Your web page}
%\thanks{}
%\altaffiliation{}
\affiliation{Department of Physics and Astronomy, University of Iowa, Iowa City, Iowa 52240, USA}
%\affiliation{Applied Optical and Plasma Sciences, Sandia National Laboratories, Albuquerque, New Mexico 87185, USA}
%\affiliation{Sandia National Laboratories, Albuquerque, New Mexico 87185, USA}
\author{Scott D. Baalrud}
\affiliation{Department of Physics and Astronomy, University of Iowa, Iowa City, Iowa 52240, USA}

\author{Matthew M. Hopkins}
\affiliation{Applied Optical and Plasma Sciences, Sandia National Laboratories, Albuquerque, New Mexico 87185, USA}

\author{Benjamin T. Yee}
\affiliation{Applied Optical and Plasma Sciences, Sandia National Laboratories, Albuquerque, New Mexico 87185, USA}

\author{Edward V. Barnat}
\affiliation{Applied Optical and Plasma Sciences, Sandia National Laboratories, Albuquerque, New Mexico 87185, USA}

% Collaboration name, if desired (requires use of superscriptaddress option in \documentclass). 
% \noaffiliation is required (may also be used with the \author command).
%\collaboration{}
%\noaffiliation

\date{\today}

\begin{abstract}
%The form of a sheath near a small electrode with changing bias, from below to above the plasma potential, is studied using 2D particle-in-cell (PIC) simulations. When the electrode is biased within $T_e/2e$ below the plasma potential, the electron velocity distribution functions (EVDFs) exhibit a loss-cone-like truncation due to fast electrons overcoming the small potential difference between the electrode and plasma. For a negative bias within $T_e/2e$ of the plasma potential no sheath is present and the plasma remains quasineutral up to the electrode. In this case the EVDF truncation leads to a presheath-like density and flow velocity gradient. When the electrode bias is increased to a few times $T_i/2e$ above the plasma potential and electron sheath is present, and all ions are repelled. Here the truncation driven behavior persists, accompanied by a shift in the maximum value of the EVDF not present in the negative bias cases. For the electron sheath case, the electrons have a flow moment at the sheath edge due to both a velocity space truncation and a shift of the maximum EVDF value. As the electrode bias is increased further, the EVDF is deformed by the sheath electric field. 

The form of a sheath near a small electrode, with bias changing from below to above the plasma potential is studied using 2D particle-in-cell (PIC) simulations. Five cases are studied: (A) an electrode biased more than the electron temperature ($T_e/e$) below the plasma potential, (B) an electrode biased less than $T_e/2e$ below the plasma potential, (C) an electrode biased nearly at the plasma potential, (D) an electrode biased more than $T_i/2e$ but less than $T_e/2e$ above the plasma potential, and (E) an electrode biased much greater than $T_e/2e$ above the plasma potential. In case (A), the electron velocity distribution function (EVDF) is observed to be Maxwellian with a Boltzmann-type exponential density decay through the ion sheath and presheath. In cases (B) and (C), the EVDFs exhibit a loss-cone type truncation due to fast electrons overcoming the small potential difference between the electrode and plasma. No sheath is present in this regime, and the plasma remains quasineutral up to the electrode. The EVDF truncation leads to a presheath-like density and flow velocity gradient. In case (D) an electron sheath is present, and essentially all ions are repelled. Here the truncation driven behavior persists, but is accompanied by a shift in the maximum value of the EVDF that is not present in the negative bias cases. In case (E), the flow shift becomes greater and the loss-cone moves further into the tail of the EVDF. In this case the flow moment has significant contributions from both the flow shift of the EVDF maximum, and the loss-cone truncation. 

\end{abstract}

% insert suggested PACS numbers in braces on next line

\maketitle %\maketitle must follow title, authors, abstract and \pacs

% Body of paper goes here. Use proper sectioning commands. 
% References should be done using the \cite, \ref, and \label commands
\section{INTRODUCTION}

Electron sheaths can form around plasma boundaries that are biased above the plasma potential. In the absence of secondary emission of electrons from the boundary, the electrode can only be biased above the plasma potential provided that the electron collecting surface is small enough that global current balance in the plasma can be maintained\cite{2007PhPl...14d2109B}. This requirement that the electron collecting area is small typically precludes this situation from happening near large boundaries, hence the typical electron sheath is found around small electrodes such as Langmuir probes collecting the electron saturation current\cite{1926PhRv...28..727M}.

The conventional wisdom has been that electron sheaths interface with the bulk plasma without a presheath \cite{1926PhRv...28..727M, 2005PhPl...12e5502H}, do not pose a significant perturbation to the bulk plasma, and do not require any significant acceleration region for electrons entering the sheath\cite{1991JPhD...24..493R,2006PSST...15..773C}. However, recent experiments and simulations at low temperature plasma conditions, \cite{2015arXiv150805971Y} along with theoretical work \cite{2015PhPl...22l3520S}, have shown that the electron sheath is accompanied by a presheath. In this presheath the electrons achieve a flow moment comparable to the electron thermal speed by the sheath edge. Unlike the ion presheath, which accelerates ions through a ballistic response to the electric field, the electron presheath accelerates electrons through a force due to a pressure gradient. Simulations of the electron presheath have demonstrated that this pressure gradient term in the electron momentum equation is in fact the dominant term\cite{2015PhPl...22l3520S}. 

The electron sheath and presheath are fundamental features of plasma boundary interactions. Although they have been rarely studied, there are potentially important applications of a deeper understanding of this fundamental phenomenon. One example is the low potential side of double layers, such as those found in the auroral upward current region\cite{2004JGRA..10912220E} and double layers in helicon plasma thrusters\cite{2007ApPhL..91t1505C}. Other areas where electron sheaths are known to be present are tethered space probes\cite{1996JGR...10117229S}, near electrodes used to induce circulation in dusty plasmas\cite{1998PhRvL..80.4189L}, as well as the sheath prior to the formation of anode spots on small and gridded electrodes \cite{1991JPhD...24.1789S, 2009PSST...18c5002B, 2008PSST...17c5006S}.      

In this paper, the pressure-gradient-driven flow that arises as the bias of a small electrode is swept from several volts below to several volts above the plasma potential is studied using 2D particle-in-cell (PIC) simulations. During the transition from ion to electron sheath the electron velocity distribution function (EVDF) develops a loss cone as the electrode bias nears the plasma potential. As the electrode bias begins to exceed the plasma potential, the EVDF acquires a significant flow shift in addition to the loss cone. The nature of the EVDF in this transition is quantified by focusing on five electrode biases. In case (A) the electrode is biased several times $T_e/2e$ below the plasma potential so that an ion sheath is present. Here $T_e/2e$ is the typical change in potential seen in ion presheaths, and is used as a measure of how weak the electrode can be biased below the plasma potential before a significant number of electrons overcome the potential barrier and are collected. In this case nearly all electrons are repelled and the EVDF remains Maxwellian. Case (B) consists of an electrode biased less than $T_e/2e$ below the plasma potential. In this case the plasma remains quasineutral up to the electrode surface. At this bias, not all electrons are repelled resulting in the formation of a loss cone, which leads to presheath-like density and flow velocity gradients that increase as the electrode is approached. In case (C), when the electrode is biased nearly at the plasma potential, the loss cone and presheath-like gradients persist. In case (D), an electron sheath is observed at biases more than $T_i/2e$ but less than $T_e/2e$ above the plasma potential. Here $T_i/2e$ is significant since below this bias a significant number of ions overcome the potential barrier and are collected by the electrode. In case (D) the EVDF continues to have a loss cone, but also has developed a significant flow shift. In case (E) the electrode is biased to a value much greater than the plasma potential. In this case the flow shift is more visible and the EVDF undergoes a distortion due to the acceleration in the strong electric field.

This paper is organized as follows. Sec. II describes the simulation setup and presents the key observations from simulations at different values of the electrode bias. Sec. III discusses the results of the simulations and explains the observations. Sec. IV gives concluding remarks.

\section{SIMULATIONS}
Simulations were performed using Aleph, an electrostatic PIC code \cite{2012CoPP...52..295T}. For the present work, a 2D triangular mesh was used to discretize the 2D domain shown in Fig. 1a. The 15 cm by 5 cm domain had three boundaries with a Dirichlet $V=0$ boundary condition and one reflecting Neumann boundary condition so that the simulation represents a physical area of 15 cm by 10 cm. An electrode of length 0.1 cm and width 0.02 cm was placed perpendicular to the center of the reflecting boundary. The time step of 0.5$\times10^{-4} \ \mu s$ resolved the electron plasma frequency and ensured that the plasma particles satisfied the CFL condition\cite{courant1928partiellen}. Each simulation ran for 800,000 time steps totaling 40 $\mu s$ of physical time. The size of each mesh element was 0.02 cm such that the electron Debye length was resolved. Particles used included helium ions and electrons, each with a macroparticle weight of 2000. The helium plasma was sourced in the volume at a rate of $10^8 cm^{-3} \ \mu s^{-1}$ at temperatures of 0.08 eV for ions and 4 eV for electrons.

\begin{table}
\begin{tabular}{l || c | c | c | r}
Simulation & $\ \ \phi_p \ \  $ & $\ \ \phi_E \ \  $ & $\ \ T_e \ \ $ & Desired criteria \ \ \ \ \ \ \ \\
& (V)& (V)& (eV)&for $\phi_E-\phi_p$ \ \ \ \ \ \ \ \ \ \\
\hline
\hline
A&2.2&-25&0.62&$\phi_E-\phi_p< -T_e/2e$ \\
B&5.7&5.5&1.55&$\ 0>\phi_E-\phi_p> -T_e/2e$\\
C&6.5&6.5&1.88& $\phi_E-\phi_p\sim 0 \ \ \ \ \ \ \ \ $\\
D&7.6&8&2.29&$T_e/2e>\phi_E-\phi_p>T_i/2e \ \ \ $\\
E&18.5&25&5.17&$\phi_E-\phi_p>T_e/2e \ \ \ $ \\
\hline
\end{tabular} 
\caption{Summary of the electrode bias $\phi_E$, plasma potential $\phi_p$, electron temperature $T_e$, and the desired value for $\phi_E-\phi_p$ in each simulation. }
\end{table}

Simulations for the five cases (A)-(E), mentioned previously in Sec. I, were carried out and results are summarized in Table I. The flow moments, densities for electrons and ions, and potential profiles are shown in Fig. 2. The figure displays values along a line perpendicular to the electrode, along the reflecting boundary. These values are from the average of the last 100,000 time steps of field data output by Aleph. Note that the non-smooth behavior in the velocity moment data is due to the velocity moment being a cell-based value, which is constant across each cell in the final output of the simulations. Since the simulations were 2D, the temperature values used here were computed from the x and y velocity components only, i.e. $T_e = n_e\int d^2v m_e(v_{r,x}^2+v_{r,y}^2)f_e/2$, where $f_e$ is the EVDF, $n_e$ is the electron density, and $v_{r,i}=(\vc{v}-\vc{V_e})\cdot\hat{\vc{i}}$. Two-dimensional velocity moments have previously been used when comparing theory to 2D simulations\cite{2015PhPl...22l3520S,2015PPCF...57d4003B}. The electron and ion flow velocities in Fig. 2 are normalized by $v_{T_e}=\sqrt{T_e/m_e}$ and $v_B=\sqrt{T_e/m_i}$ respectively. Here the 2D values of $T_e$ were computed approximately 0.5 cm from the electrode surface, and were 0.62 eV, 1.55 eV, 1.88 eV, 2.29 eV, and 5.17eV for cases (A)-(E) respectively.

Using data from simulations the terms of the momentum equation, 
\begin{equation}
m_en_e\vc{V}_e\cdot\nabla\vc{V}_e=-ne\vc{E}-\nabla\cdot\mathcal{P}_e-\vc{R}_e,
\end{equation}
were evaluated. Here $m_e$, $n_e$, and $\vc{V_e}$ are the electron mass, density, and flow moment, $\mathcal{P}_e$ is the pressure tensor defined below, and $\vc{R}_e$ is a friction term. The $\hat{x}$ vector components $m_en_e\vc{V}_e\cdot\nabla\vc{V}_e\cdot\hat{x}$ and $\nabla\cdot\mathcal{P}_e\cdot\hat{x}$ of the terms of the momentum equation were computed from particle location and velocity data at 20 different time slices separated by $5\times10^{-3}\mu s$ each. The velocity moments of the electron velocity distribution function $f_e(\vc{v})$ used above are 
\begin{equation}
\vc{V}_e=\frac{1}{n_e}\int d^2v \vc{v} f_e(\vc{v}),
\end{equation}
\begin{equation}
n_e=\int d^2v  f_e(\vc{v}),
\end{equation}
and
\begin{equation}
\mathcal{P}_e=m_e\int d^2v (\vc{v}-\vc{V_e})(\vc{v}-\vc{V_e})f_e(\vc{v})=p_e\mathcal{I}+\Pi_e,
\end{equation}
where the pressure tensor is decomposed into the scalar pressure $p_e$ multiplied by the identity tensor $\mathcal{I}$ and stress tensor $\Pi_e$. Based on observation, it was assumed that the dominant gradients in $\vc{V_e}$ and $\mathcal{P}_e$ were in the $\hat{x}$ direction. These moments along with the value of $-ne\vc{E}\cdot\hat{x}$ from the time averaged mesh based data are also shown in Fig. 2. Once again, all values were computed in 2D using only x and y velocity information.

Using particle location and velocity data from the same 20 time slices mentioned above, the EVDFs were obtained by plotting the 2D histogram of the x and y velocity components for particles. First, the distributions were obtained in three 0.1 by 0.1 cm boxes for all five simulations, the locations of which are indicated by the black boxes in Fig. 1b. EVDFs for all simulations are shown in Fig. 4. For the case of the electrode biased at 25V additional EVDFs, shown in Fig. 4, were obtained in the region indicated by the small red boxes in Fig. 1b to better resolve the EVDF within the sheath. Note that a slight asymmetry is present in the y velocity components in Fig. 3, this is due to the averaging box not being aligned with the symmetry axis as shown in Fig. 1b. This asymmetry is less visible in Fig. 4 due to the smaller averaging area. The interpretation of the data for these five simulations are discussed in the following section.

\

\section{DISCUSSION}

In the previous section, simulations with five different values of electrode bias were presented: strong ion and electron sheath cases (A) and (E), and three cases (B)-(D) with the electrode biased near the plasma potential. In this section, flow moments and EVDFs calculated from these simulations are used to study the development of pressure gradient driven electron flow in the transition from ion to electron sheath. 

%Here, the flow is shown to develop from an initial loss-cone truncation in the EVDF when the electrode is biased slightly below the plasma potential in simulation (B). In this case, the plasma remains quasineutral up to the electrode, a result that has not been previously observed. Once the electrode bias exceeds the plasma potential the EVDF also develops a flow shift, which has been observed in simulations of the electron presheath\cite{2015PhPl...22l3520S,2015arXiv150805971Y}. 

First, the behavior of electrons near the strong ion sheath case (A) is examined. The fluid moments and fields for this case are shown in Fig. 2 and the EVDFs in Fig. 3. In the ion presheath and sheath the electrons have no flow, except for a slight flow shift directed out of the sheath within 0.2cm of the electrode, which is due to the expulsion of a small number of electrons sourced in the sheath. Although the flow is non-zero within 0.2 cm, the density is small enough that $n_e\vc{V_e}$ is negligible.  The EVDFs show no truncation or flow shift, which corresponds to the expectation based on a Boltzmann density profile $n_e\sim \exp(-e\Delta\phi/T_e)$ in the presheath. This expectation is verified in Fig. 5 using the computed 2D electron temperature and simulated values of $\Delta\phi$. The EVDFs demonstrate that as the electrode is approached the density drops off rapidly, with only a few particles present within 0.1 cm of the electrode surface. A Boltzmann density profile is expected when the EVDF is Maxwellian, which can be seen from the electron momentum equation given in Eq. (1). In this cassette friction and flow terms can be neglected, the pressure tensor reduces to $n_eT_e\mathcal{I}$, and its gradient is balanced by the electric field, leading to the exponential relation mentioned above. While other distributions might satisfy the zero flow moment requirement, the Maxwellian supports an exponential decay for particles that follow characteristics defined by the electric field\cite{1873Natur...8..537C}.

When the electrode was biased within $T_e/2e$ below the plasma potential, as it is in case (B), the assumptions of the Boltzmann relation for the density profile break down. At such a small bias a significant number of electrons have energy greater than $T_e/2$ and reach the electrode, resulting in a truncation of the EVDF. The only electrons directed away from the electrode near it's surface are those sourced in this region. However, this population is small and essentially negligible. The effects of this truncation can be seen in Fig. 3. Near the surface of the electrode, the truncated part of the EVDF forms a loss-cone type distribution with an angle that narrows when moving further into the bulk plasma. As shown in Fig. 6, there is an area that is geometrically inaccessible to electrons moving from left to right, forming a shadow. In this region fast electrons directed towards the electrode are collected, and no electrons at high energy are available to fill in the part of the distribution directed away from the electrode. In the case of a strong ion sheath this does not occur, electrons directed towards the electrode are reflected and the distribution remains Maxwellian.  

In the cases (B) and (C) the difference between the plasma potential and electrode are $0>\phi_E-\phi_p> -T_e/2e$ and $\phi_E-\phi_p\sim 0$, as can be seen in the potential plotted in Fig. 2. The density profiles for ions and electrons, along with the small potential gradients, indicate that in these cases the plasma remains quasineutral up to the electrode. Although there is no breakdown of quasineutrality, there are still significant flow moments and density gradients for electrons and ions. Unlike the typical ion presheath, the ions only achieve a flow velocity of $\sim0.4v_B$ by the electrode for the case slightly below the plasma potential, while the electrons have a flow of $\sim0.6v_{T_e}$, with both flows directed towards the electrode. The electric field term for case (B) is weaker than the pressure term, and for case (C) the field term is negligible, suggesting the flows in this regime are not related to the electric field. Gradients in the electron densities and flow velocities can be explained by the widening of the loss cone. As the electrode is approached the loss cone angle increases while there is little or no shift in velocity of the maximum value of the EVDF in Fig. 3. The increasing flow moment is a result of the increasing angle of the loss cone. The continuity equation $\nabla\cdot(n_e\vc{V_e})=0$ requires that an increase in flow moment must be accompanied by a decrease in density. The computed velocity moment gradients in Fig. 2 indicate that the pressure tensor gradient, due to the increasing truncation, plays an important role when the electrode is biased near the plasma potential.  

In case (D) the behavior changes. The most notable difference is that the plasma no longer remains quasineutral up to the electrode, but a well defined electron sheath has formed. When $\phi_E-\phi_p\sim T_e/2e$ nearly all of the ions are repelled due to the fact that $T_e\gg T_i$. This can be seen in the ion density profile, which shows very few ions in the electron sheath. Here, the electrons achieve a flow velocity of $\sim0.6\sqrt{T_e/m_e}$ by the sheath edge, which is slightly slower than the previously predicted value of $\sqrt{T_e/m_e}$\cite{2015PhPl...22l3520S}. However, the exact sheath edge position is difficult to determine in time averaged data due to streaming instability driven fluctuations which have been discussed previously\cite{2015PhPl...22l3520S}. In addition, the form of the EVDF, including the extent of the loss cone, would need to be accounted for in the calculation of the sheath edge speed for electrons. Nevertheless, the flow of electrons into the sheath can be seen to be driven by pressure tensor gradients in Fig. 2, with some portion of the flow due to the shift in the maximum value of the EVDF, and the rest due to the truncation visible in Fig. 3. 

The presheath behavior remains qualitatively similar as the electrode bias is further increased, as demonstrated by case (E). The presheath behavior is largely the same as that in case (D), a shift in maximum value of the EVDF and a loss cone is present near the sheath as shown in Fig. 3. The main difference between simulation (E) and simulation (D) is the deformation of the EVDF within the sheath, which is shown in detail in Fig. 4. Near the sheath edge, at x=0.165 cm,  the EVDF resembles the loss cone shaped distributions of Fig. 3. Moving into the sheath, the EVDF becomes stretched out, with electrode-directed electrons becoming more and more depleted due to the increasing strength of the electric field. A similar deformation of the EVDF has been seen in Vlasov simulations of an electron sheath around a cylindrical probe\cite{2013PhPl...20a3504S}.

The electron presheath behavior in simulations (D) and (E) were slightly different than that previously described in the 1D model of ref. [7], however, it also shares some of the same qualities. Previous analysis was based on the 1D projections of the EVDF which look largely Maxwellian in the direction perpendicular to the electrode surface. Based on the observation of 1D EVDFs the electron presheath was modeled describing the electrons with a flow-shifted Maxwellian and ions with a Boltzmann density profile. Using these assumptions the dominant term in the electron momentum equation, Eq. (1), is the pressure gradient, which was exactly $T_e/T_i$ larger than contribution of the presheath electric field. This model proposed a significant deviation from the conventional assumption of a half-Maxwellian at the sheath edge\cite{1961JAP....32.2512M,1962JAP....33.3094M,2005PhPl...12e5502H}. It also showed that the electron sheath has a presheath that causes a significant perturbation far into the plasma. The results in this paper present a refined picture compared to the previous 1D model. Like the 1D model, a long presheath was visible in the electron flow moments shown for simulations (D) and (E) in Fig. 2. The EVDFs discussed above demonstrate that, for the electron sheath cases, the flow velocity is due to both a loss-cone like truncation and a flow shift. The truncation due to the loss cone leads to a situation where the electron flow velocity is heavily influenced by gradients in the pressure tensor. These are treated using a scalar pressure gradient in the 1D model. When the electrode is biased at or slightly below the plasma potential the EVDFs have a loss-cone like truncation, but with no flow shift, demonstrating that the flow shift is present only when the electrode is biased sufficiently positive that an electron sheath forms.

\section{CONCLUSION}

In this paper, the physics of the sheath and presheath was explored using electron velocity data from 2D PIC simulations. When the electrode was biased less than $T_e/2e$ below the plasma potential, as in case (B), the plasma remained quasineutral up to the electrode surface, but still had presheath-like density and flow velocity gradients for both electrons and ions. Using EVDFs obtained from individual electron velocity data, the flow and density gradients were shown to originate from a loss-cone-like truncation of the EVDF due to fast electrons overcoming the $T_e/2e$ potential barrier. This was described as an effect of the breakdown of the assumptions behind the Boltzmann density relation, that arises due to a gradient in the stress tensor.  

In case (D), when the electrode was biased less than $T_e/2e$ and greater than $T_i/2e$ above the plasma potential, a well defined electron sheath was present. In this case, since $T_e\gg T_i$, ions were observed to be completely repelled from the electrode. When an electron sheath was present, the EVDFs had flow velocity moments with contributions from both a flow shift, and a loss-cone-like truncation. As the electrode was approached from the bulk plasma, the loss cone angle was observed to widen, increasing the flow moment and leading to presheath behavior driven by a pressure tensor gradient. The onset of this behavior occurs in the ion sheath case when the electrode bias is sufficiently close to the plasma potential that electrons are not completely repelled, leading to an addition to the pressure tensor gradient term in Eq. (1). This contribution is due to the increasing value of the stress tensor, in Eq. (4), which grows in magnitude as the electrode is approached and the EVDF truncation becomes more pronounced. 

Finally, the electron presheath behavior was compared with the 1D model described previously. In both the present simulations, and the 1D model, a long electron presheath is present in which electrons are accelerated close to their thermal speed by forces associated with a gradient in the pressure. The 2D simulations reveal a loss-cone nature, in addition to the flow shift, that is absent in the 1D model. The 2D simulations observe contributions to the force driving flow from both the scalar pressure gradient and the stress tensor gradient, whereas a 1D model accounts only for the scalar pressure contribution.
\section*{Acknowledgments}
The first author would like to thank James Franek and Andrew Fierro for reading the manuscript. This research was supported by the Office of Fusion Energy Science at the U.S. Department of Energy under contract DE-AC04-94SL85000. The first author was also supported by the U.S. Department of Energy, Office of Science, Office of Workforce Development for Teachers and Scientists, Office of Science Graduate Student Research (SCGSR) program. The SCGSR program is administered by the Oak Ridge Institute for Science and Education for the DOE under contract number DE-AC05-06OR23100.

% Create the reference section using BibTeX:

%\nocite{*}
\bibliography{nosheath}

\

\newpage

\

\newpage

\onecolumngrid

%\begin{figure}
%\includegraphics[scale=.5]{domain_image.pdf}
%\caption{The simulation domain with a color map indicating the ion density. The electrode is placed perpendicular to the upper reflecting boundary and position along the horizontal axis is indicated by the dashed line. }
%\end{figure}
\begin{figure}
\includegraphics[scale=.5]{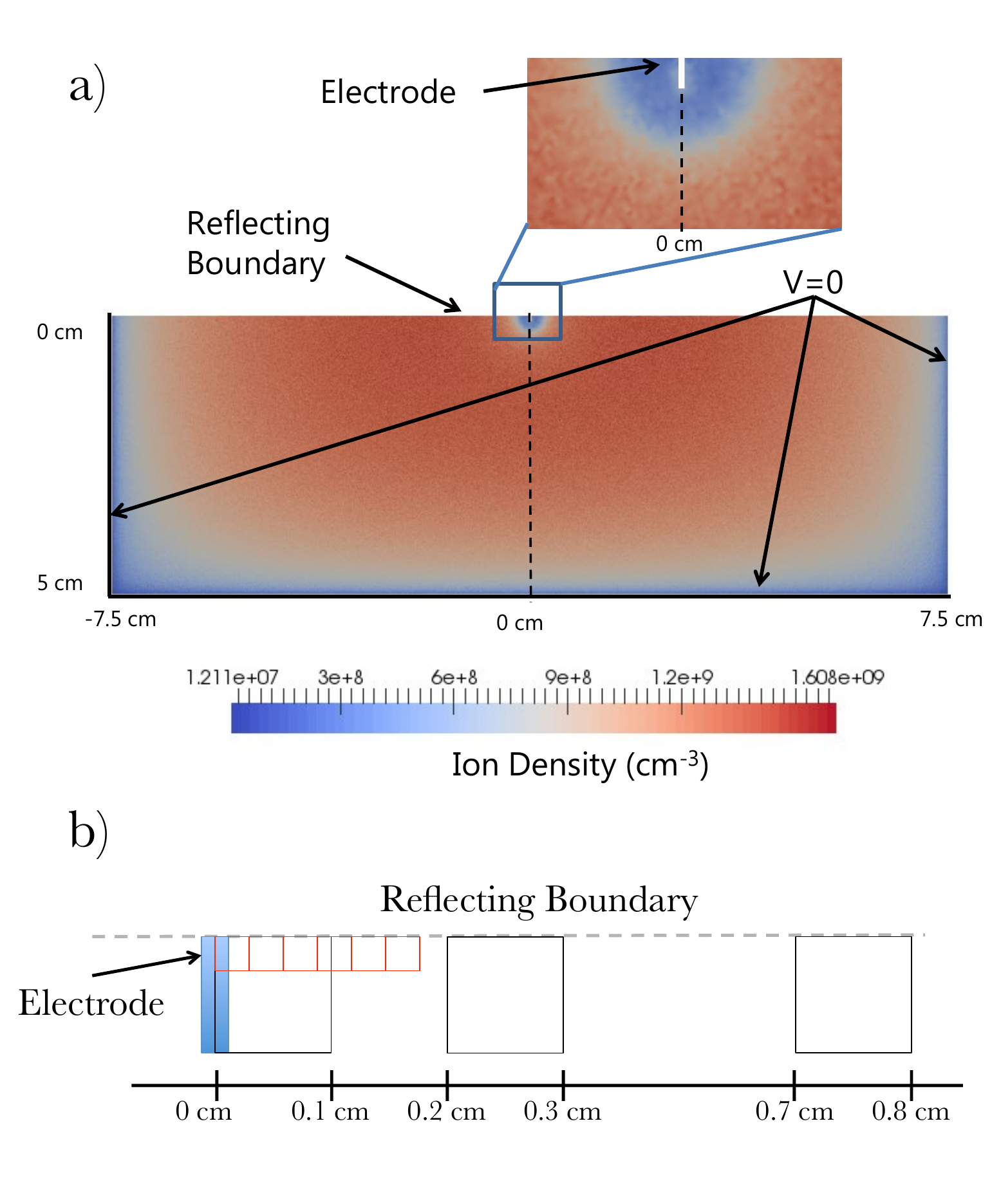}
\caption{a) The simulation domain with a color map indicating the ion density. The electrode is placed perpendicular to the upper reflecting boundary and position along the horizontal axis is indicated by the dashed line. b) Boxes displaying the limiting areas used for the calculation of EVDFs. The large black boxes were used for Fig. 3, while the small red boxes were used for Fig. 4.}
\end{figure}

\begin{figure}[ht]
  \centering
  \savebox{\myimage}{\includegraphics[scale=.92]{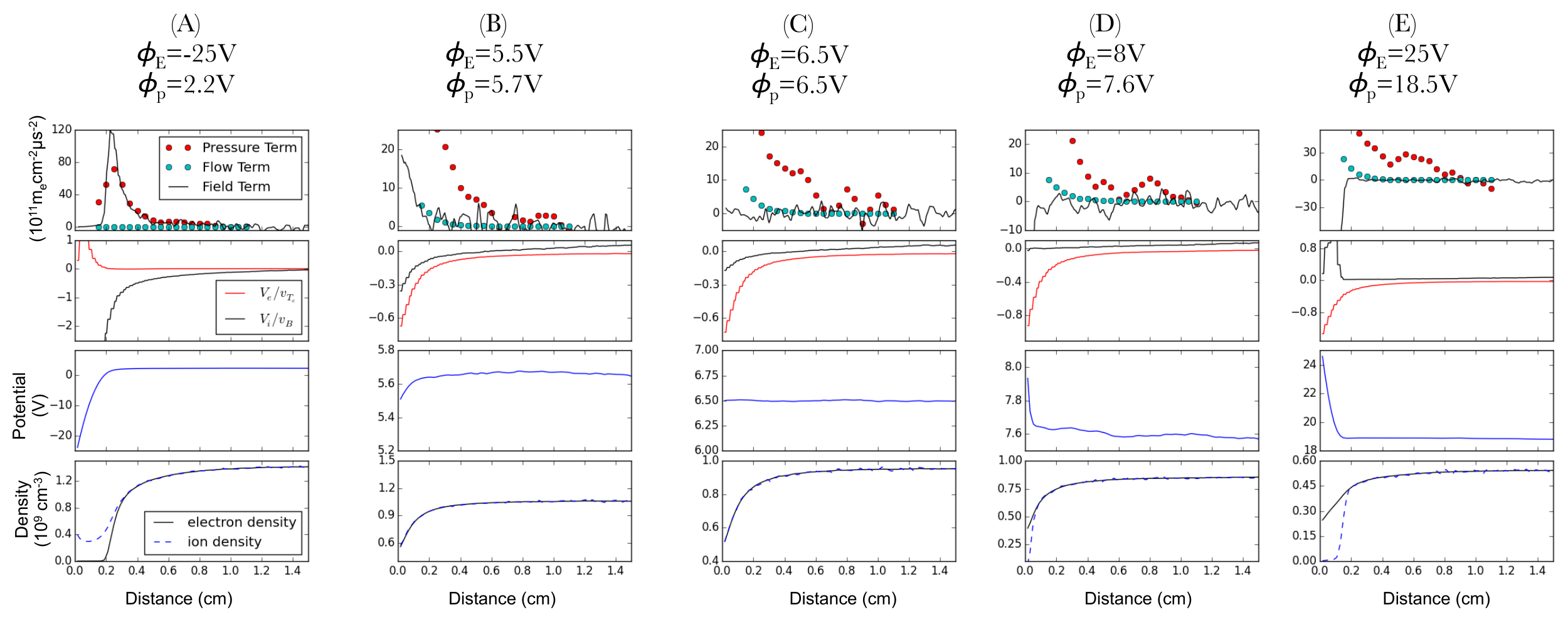}}%{\rule{100pt}{150pt}}% Image to be included
  \rotatebox{90}{% Rotate 90 CCW
    \begin{minipage}{\wd\myimage}
      \usebox{\myimage}
      \caption{Computed terms of the momentum equation, flow moments, potential, and density profiles for simulations (A)-(E). All quantities were calculated from mesh based values except for the pressure and flow terms of the momentum equation which were computed from particle positions and velocities as described in the text. In the top panel of each plot field term refers to $-ne\vc{E}\cdot\hat{x}$, flow term refers to $m_en_e\vc{V}_e\cdot\nabla\vc{V}_e\cdot\hat{x}$, and the pressure term is $\nabla\cdot\mathcal{P}_e\cdot\hat{x}$. Note that the non-smooth behavior in the velocity moment data is due to this being a cell based quantity, which is output as constant across each cell. }
    \end{minipage}}
\end{figure}

%\begin{figure}
%\includegraphics[scale=.4]{Figs_final/11.pdf}
%\caption{ Boxes displaying the limiting areas used for the calculation of EVDFs. The large black boxes were used for Fig. 4, while the small red boxes were used for Fig. 6.}
%\end{figure}

\begin{figure}[ht]
  \centering
  \savebox{\myimage}{\includegraphics[scale=.9]{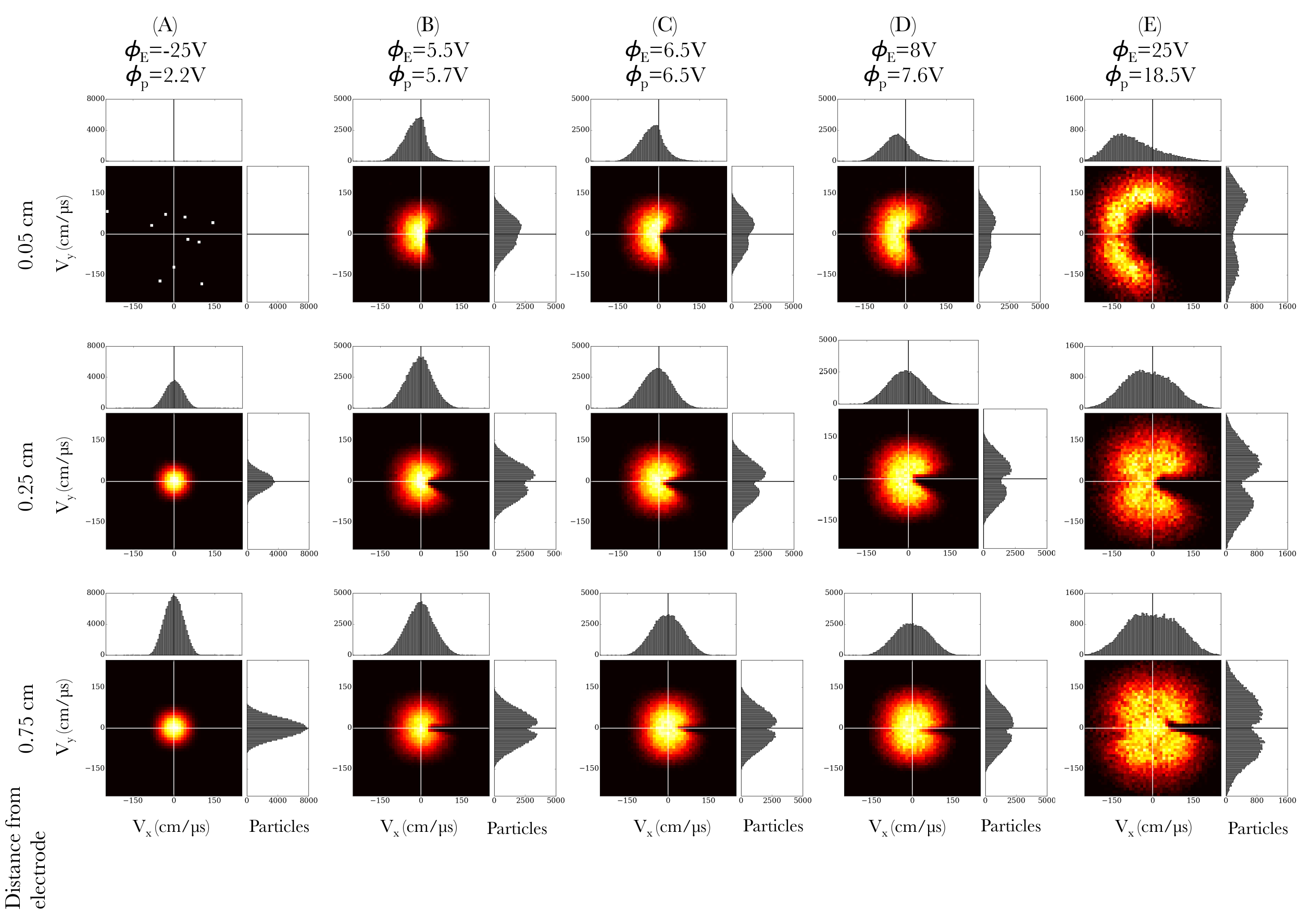}}%{\rule{100pt}{150pt}}% Image to be included
  \rotatebox{90}{% Rotate 90 CCW
    \begin{minipage}{\wd\myimage}
      \usebox{\myimage}
      \caption{EVDFs for simulations (A)-(E). The EVDF locations are marked as the centers of the black boxes shown in Fig. 1b. The EVDF on the top left is next to the electrode and few electrons are present. Note that the slight asymmetry present in the y velocity components is due to the averaging box being off-center of the symmetry axis marked by the reflecting boundary in Fig. 1b.}
    \end{minipage}}
\end{figure}

\begin{figure}
\includegraphics[scale=1]{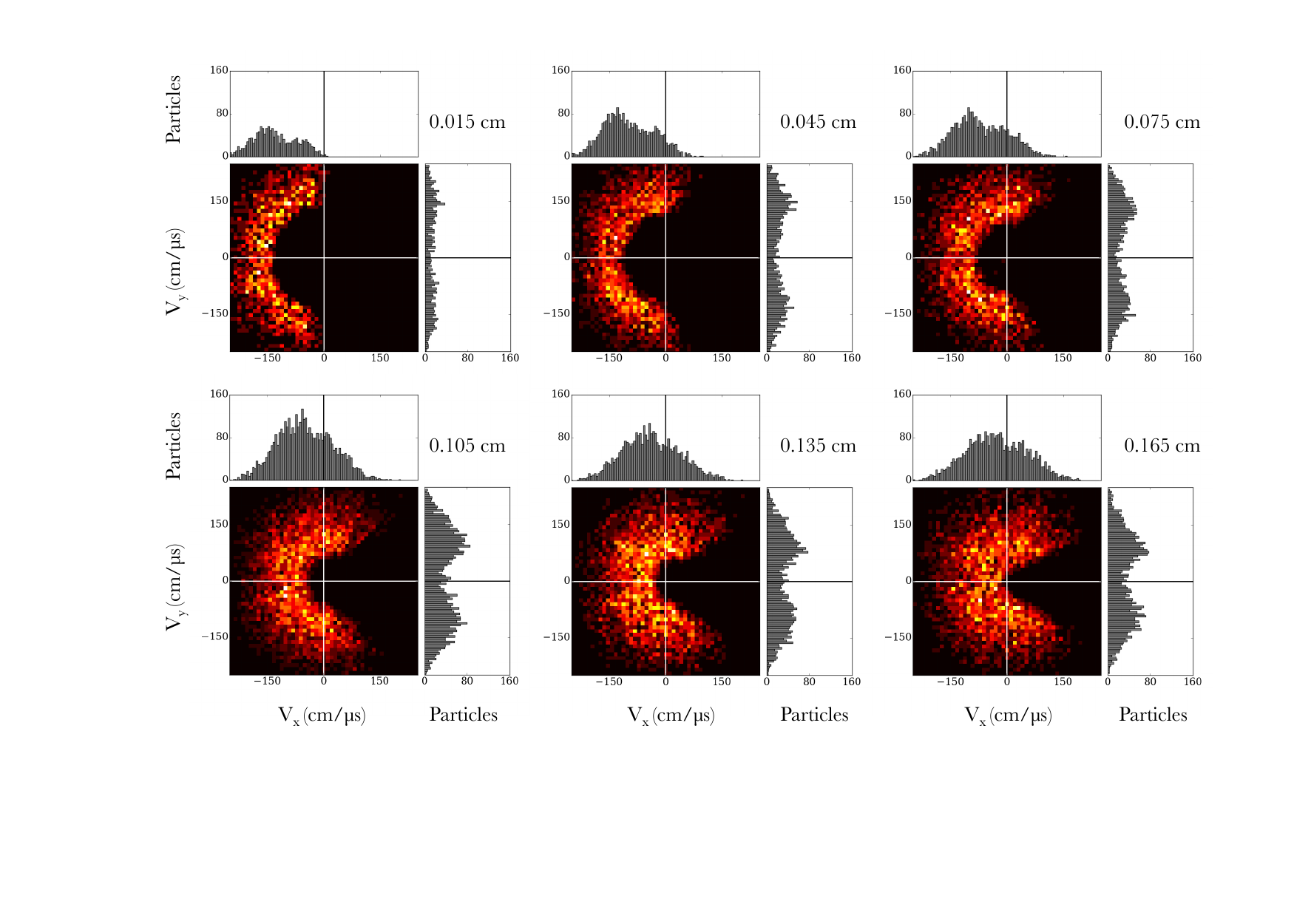}
\caption{EVDFs for simulation (E), the case of a strong electron sheath. The EVDF locations are marked as the centers of the red boxes shown in Fig. 1b. Finer spatial resolution plots are shown to demonstrate the deformation of the EVDF within the electron sheath. }
\end{figure}

\

\begin{figure}[h!]
\includegraphics[scale=.35]{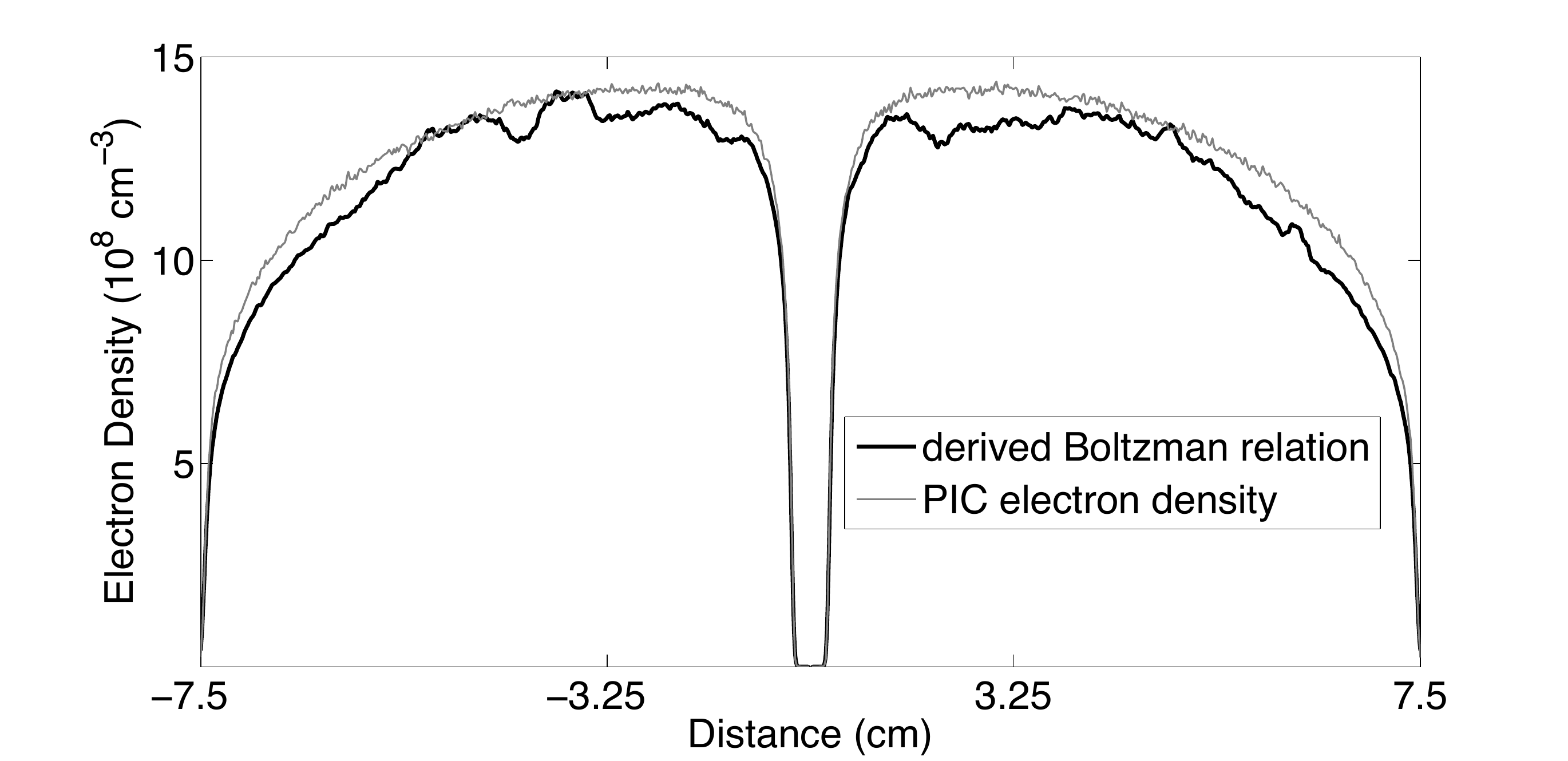}
\caption{ Comparison of the electron density from simulation (A) with the Boltzmann relation using simulated values of the potential and the 2D temperature. }
\end{figure}

\begin{figure}[h!]
\includegraphics[scale=.5]{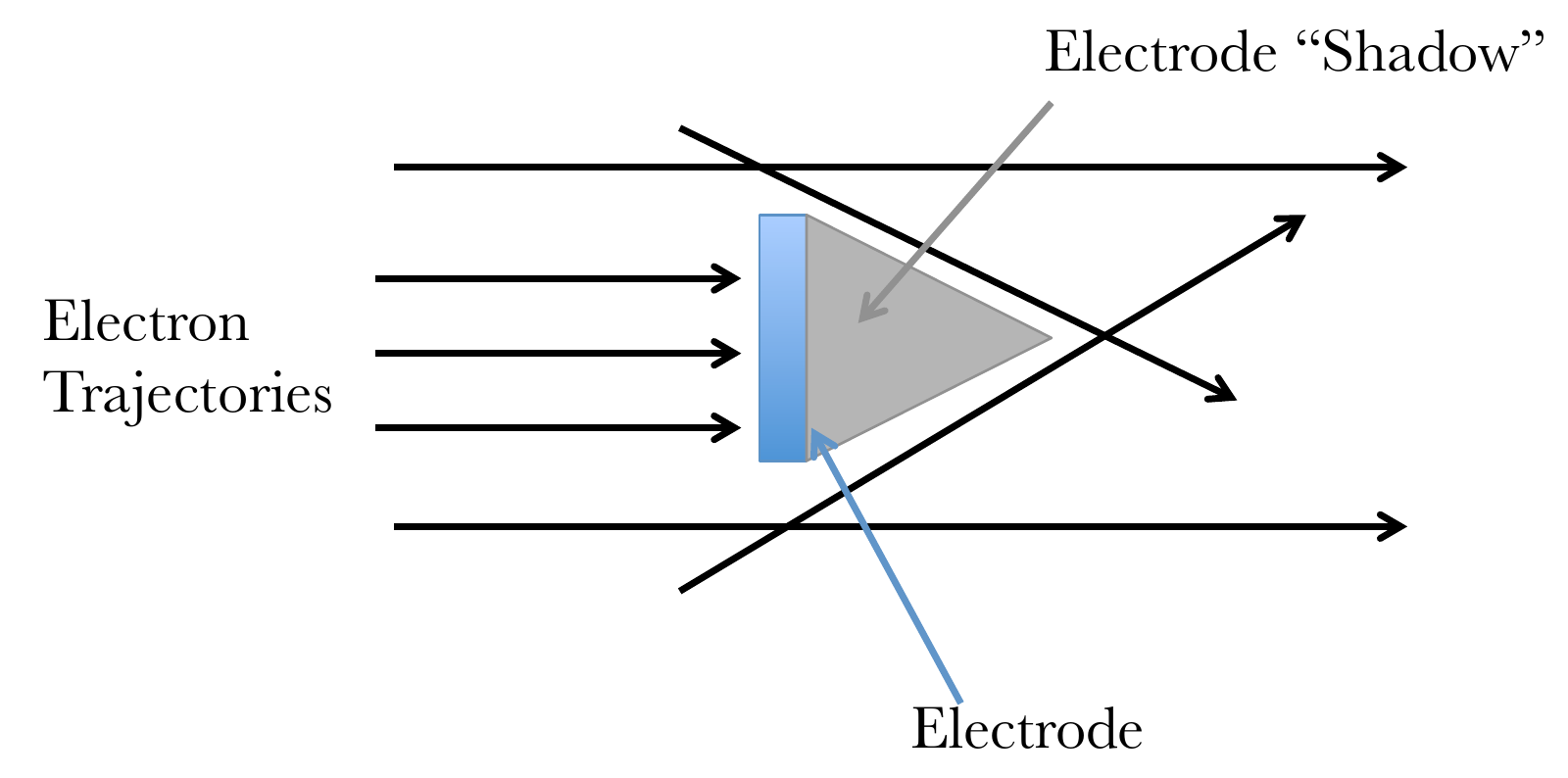}
\caption{ Electron trajectories for right moving electrons are shown to demonstrate how a boundary can cause a truncation in a velocity distribution function, such as the loss-cone-like truncation.}
\end{figure}

\end{document}